\documentclass[%
 reprint,
superscriptaddress,
 amsmath,amssymb,
 aps,
pra,
longbibliography,
]{revtex4-2}

\usepackage{graphicx}
\usepackage{dcolumn}
\usepackage{bm}

\usepackage{blindtext}
\usepackage{siunitx}

\usepackage{lineno}

\begin{document}

\preprint{APS/123-QED}

\title{Secondary electron yield from aluminium-coated foils for muon tagging and beam monitoring up to 60 MeV/c}
\author{Gianluca Janka}
\email{gianluca.janka@psi.ch}
\affiliation{PSI Center for Neutron and Muon Sciences CNM, 5232 Villigen PSI, Switzerland}
\author{Michael W. Heiss}
\author{Maxime Lamotte}
\author{Xiao Zhao}
\author{Thomas Prokscha}
\email{thomas.prokscha@psi.ch}
\affiliation{PSI Center for Neutron and Muon Sciences CNM, 5232 Villigen PSI, Switzerland}
\date{\today}

\begin{abstract}
The feasibility of foil-based muon tagging is investigated in the momentum range below \SI{60}{\mega\electronvolt\per c}, with particular focus on its applicability to the low-momentum range spanning approximately \SI{2.5}{\mega\electronvolt\per c} to \SI{20}{\mega\electronvolt\per c}, where no efficient and minimally invasive detection scheme is currently established for continuous beams. 
Secondary electron emission from a \SI{7}{\micro\meter} Mylar foil coated with \SI{50}{\nano\meter} aluminium is investigated using a continuous negative muon beam with nominal momenta between \SI{12}{\mega\electronvolt\per c} and \SI{60}{\mega\electronvolt\per c} at the $\pi$E1 beamline at PSI. The emitted electrons are detected with position-sensitive microchannel plate detectors, enabling particle tagging and spatial characterization of the beam.
The detection efficiency and corresponding secondary electron yield are extracted and benchmarked against literature data for protons, showing good agreement and confirming reliable muon tagging. The observed trend for negative muons is consistent with the well-established increase in ion-induced secondary electron emission toward lower particle velocities, suggesting improved performance in the low-momentum regime.
A proof-of-principle reconstruction of the muon beam profile is demonstrated by correlating detected electron positions with their emission point at the foil. These results establish foil-based tagging as a viable approach for combined timing and minimally invasive beam monitoring, bridging the gap between high- and low-energy-muon instrumentation.
\end{abstract}

\maketitle


\section{\label{sec:Intro}Introduction}

At continuous muon beamline facilities, such as the Swiss Muon Source (S$\mu$S) at PSI, Switzerland, entrance detectors based on thin scintillators with typical thicknesses of the order of \SI{100}{\micro\meter} are routinely employed. These detectors register traversing muons and provide the start time for time-resolved measurements, and are well suited for muon momenta of \SI{20}{\mega\electronvolt\per c} (corresponding to a kinetic energy of approximately \SI{2}{\mega\electronvolt}) and above. At substantially lower muon energies, such as at the Low-Energy Muon (LEM) beamline where muon energies fall below \SI{20}{\kilo\electronvolt} \cite{2008_Prokscha, 2001_Prokscha,2004_Morenzoni}, such scintillators cannot be used, as the muons would be fully stopped within the material. In this low-energy regime, an ultrathin carbon foil (\SI{10}{\nano\meter}) is instead employed: the passage of a muon liberates secondary electrons from the foil, which are subsequently detected to generate the timing signal \cite{2015_Khaw,2024_Janka}.

Between these two regimes, however, no efficient and minimally invasive particle tagging method for continuous muon beams has yet been established. Muons in this regime are of particular interest for applications such as muon spin relaxation ($\mu$SR, \cite{2022_Blundell,2024_Amato}) studies or elemental analysis via muon-induced X-ray emission (MIXE, \cite{2022_Biswas,2023_Gerchow_GIANT}), which enable the investigation of material properties at depths of hundreds of \si{\nano\meter} to \si{\micro\meter} scale. The need for improved detector concepts in this context is also reflected in the development of next-generation muon beam facilities such as the High-Intensity Muon Beams (HIMB) project at PSI, where it is noted that such beam energies will require the development of new muon entrance detectors~\cite{himb}.

In addition to timing, spatial information on the muon beam is crucial for beam diagnostics and tracking applications. Any detector placed in the beam path, however, inevitably introduces material that leads to energy loss and multiple scattering, which can degrade the beam quality and limit the achievable resolution. At present, no minimally invasive beam monitoring devices are routinely available for these beamlines. Recent developments using pixelated semiconductor detectors have demonstrated beam imaging capabilities by combining tracking of the incoming muon with tracking of its decay product \cite{2026_Mandok_muSIP}. In such setups, the muon trajectory is reconstructed using several upstream detector layers, while the decay positron is tracked in surrounding detectors and extrapolated back to its origin, allowing the determination of the muon stopping position. This approach is effective for sufficiently energetic muons that can traverse the detector layers with limited scattering. However, the required detector thicknesses, typically on the order of \SI{100}{\micro\meter}, introduce significant energy loss and multiple scattering, making this method unsuitable at lower muon momenta. In addition, the reconstruction relies on tracking both the incoming muon and its decay products. A minimally invasive, direct measurement of the muon position at the point of interaction would relax these requirements and enable operation at lower energies. Alternative approaches based on gaseous detectors are being explored, for example in the context of MIXE experiments \cite{2025_Garcia, 2025_Zhao}, but similarly introduce a non-negligible material budget along the beam path, ultimately restricting their use in the low-momentum regime where beam perturbations must be minimized.

\begin{figure*}[t!]
\centering
\includegraphics[width=2\columnwidth, trim={0 0 0 0},clip]{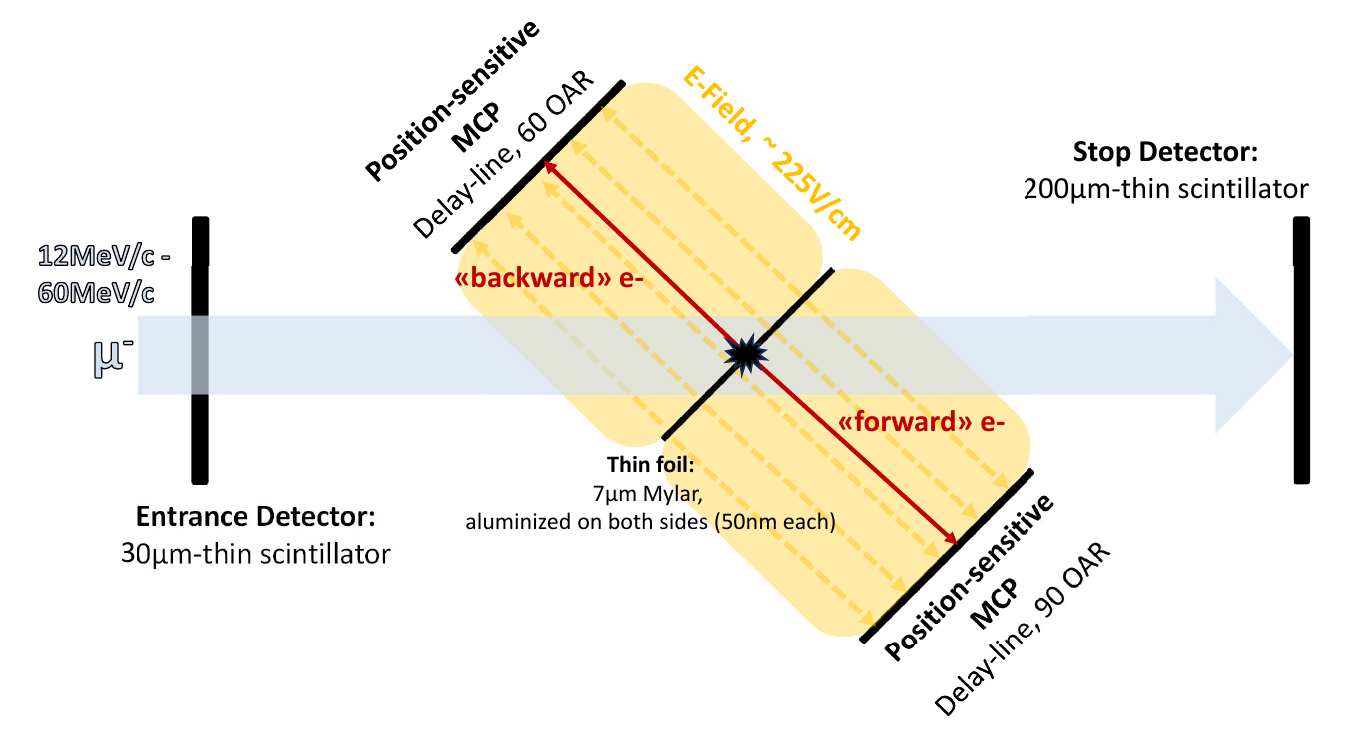}
\caption{\label{fig:setup} Schematic of the experimental setup. A continuous muon beam traverses a \SI{7}{\micro\meter} Mylar foil coated on both sides with \SI{50}{\nano\meter} aluminium, where secondary electrons are emitted. The foil is tilted by $45^\circ$ and biased to create an electric field that guides emitted electrons toward two position-sensitive MCP detectors placed upstream and downstream, detecting backward- and forward-emitted electrons, respectively. Thin plastic scintillators located before and after the foil define traversing muons.}
\end{figure*}

Secondary electron emission from thin foils provides a promising approach to address both challenges. While the secondary electron yield (SEY) from ultrathin carbon foils \cite{2024_Janka, 1962_Large_CReview,1988_Kuroki_CReview,1989_Clouvas_CReview,1997_Gelfort_CReview,1997_Vidovic_CReview,1998_Ritzau_CReview,2011_Ogawa_CReview} and other materials, such as aluminium \cite{1939_Hill_AlReview,1954_Aarset_AlReview,1974_Foti_AlReview,1977_Thornton_AlReview,1997_Castanenda_AlO2_AlReview,1988_Borovsky,1981_Hasselkamp_AlReview,1981_Koyama_AlReview,1981_Svensson_AlReview,1995_Benka_AlReview, 1979_Baragiola_AlReview}, has been extensively studied using protons and heavier ions, corresponding measurements for muons remain scarce \cite{2024_Janka, 2025_Janka_CReview}. Existing data for protons indicate substantial SEY at particle velocities corresponding to the intermediate-energy muon regime, suggesting that foil-based tagging may be feasible while also offering a path toward minimally invasive beam monitoring by transporting the emitted secondary electrons onto a position-sensitive detector, where their spatial distribution can be correlated with the muon impact position at the foil.

In this work, we measure the secondary electron yield from an aluminium-coated Mylar foil for muons with nominal momenta between \SI{12}{\mega\electronvolt\per c} and \SI{60}{\mega\electronvolt\per c}. We demonstrate that the observed yield is sufficient for particle tagging in this regime and present a proof-of-principle reconstruction of the beam spot using position-sensitive microchannel plates (MCP), highlighting the potential for combined tagging and beam monitoring. By benchmarking our results against literature data, we further show that the secondary electron yield is expected to increase with lower muon momenta, indicating that this approach should become even more effective in the low-energy regime, thereby bridging the gap between existing high- and low-energy muon tagging techniques.

\section{\label{sec:Setup}Experimental Setup}

The experiment was performed at the $\pi$E1 beamline at PSI, which provides high-intensity continuous pion and muon beams~\cite{PIE1_web, 2021_Grillenberger}. For the present measurements, negative muons with momenta between \SI{12}{\mega\electronvolt\per c} and \SI{60}{\mega\electronvolt\per c} were selected, as lower beam momenta do not provide sufficient transmission through the experimental setup. The electrostatic separator of the \mbox{$\pi$E1} beamline strongly suppresses beam contamination, which is therefore expected to be negligible for the present measurements. Muon rates ranging from a few to around \SI{100}{\kilo\hertz}, depending on beamline settings, are typical for this beamline~\cite{2023_Gerchow_GIANT}. Beam profile measurements at the $\pi$E1 beamline indicate spot sizes on the order of \SI{10}{\milli\meter} (rms), depending on the beam optics and momentum settings~\cite{2023_Gerchow_GIANT,2025_Hu}. 
The measurements were carried out under high-vacuum conditions at pressures of a few \SI{e-7}{mbar}, ensuring negligible scattering of both the muon beam and the emitted secondary electrons in the residual gas.

A schematic of the experimental setup is shown in Fig.~\ref{fig:setup}. The central element of the setup is a \SI{7}{\micro\meter}-thick Mylar foil, coated on both sides with \SI{50}{\nano\meter} aluminium, which serves as the secondary electron emission target. Secondary electrons emitted from the foil are detected by two position-sensitive MCP detectors (RoentDek,  DLD40 delay-line readout \cite{2002_Jagutzki_Roentdek}), which provide spatial and timing information of the emitted electrons. The MCPs are positioned symmetrically at a distance of \SI{67}{mm} on either side of the foil.

The foil is rotated by $45^\circ$ around the vertical axis, allowing the MCPs to be positioned outside the direct muon beam while maintaining a sufficiently uniform electric field for electron transport in the central region. The foil is biased to \SI{-4.2}{\kilo\volt}, while the front surfaces of both MCPs are held at \SI{-2.7}{\kilo\volt}, resulting in an electric field of approximately \SI{225}{\volt}/\si{\centi\meter}. The emitted electrons are guided from the foil to the MCPs by this field. The electron transport and electric field configuration were studied using SIMION simulations \cite{Simion}, providing a qualitative understanding of the electron trajectories and spatial mapping between foil and MCP positions.

The upstream MCP, facing the incoming beam, detects electrons emitted in the backward direction (named backward MCP), while the downstream MCP records forward-emitted electrons (named forward MCP). The backward MCP is of Z-stack type with an open-area ratio (OAR) of 60\%, while the forward MCP is of chevron type with an OAR of 90\%. The effective single-electron detection efficiency of the MCPs, determined from dedicated measurements at the LEM beamline, is 0.54(2) for the upstream detector and 0.78(3) for the downstream detector, respectively. These values are in good agreement with the reported absolute detection efficiencies for MCPs of the same type~\cite{2018_Fehre_Roentdek}. Both MCPs provided an active diameter of \SI{40}{\milli\meter}, defining the effective geometrical acceptance for transported secondary electrons.

A detection of a traversing muon at the foil is defined by a signal in at least one of the MCP detectors arising from secondary electrons emitted from the foil.
To define the muon beam and provide a time reference for traversing particles, two plastic scintillators are installed upstream and downstream of the foil. The entrance detector is a \SI{30}{\micro\meter}-thick scintillator with an active area of $27 \times 27$~\si{\milli\meter\squared}, while the downstream detector is a \SI{200}{\micro\meter}-thick scintillator with an active area of $25 \times 25$~\si{\milli\meter\squared}. The distance between the two scintillators is \SI{670}{\milli\meter}. While suitable for the momentum range investigated here, the entrance scintillator constitutes the dominant material budget of the setup and ultimately limits the accessible momentum range.

The detector signals were processed using the GIANT data acquisition (DAQ) system of the MIXE experiment~\cite{2023_Gerchow_GIANT}. The MCP and scintillator signals were digitized using a 16-channel SIS3316 14-bit, 250~MS/s VME module (Struck Innovative Systeme). The system provides nanosecond-scale timing resolution, sufficient to resolve coincidence signals between detectors. High voltages were supplied and controlled using ISEG NHR 42-60r modules.

\section{\label{sec:Analysis}Data Analysis}

The data consist of time-stamped digitized signals from all detector channels acquired in a triggerless continuous-stream mode. Temporal correlations between detectors are established offline by associating each signal with the nearest signal in a reference detector (here the entrance scintillator) and computing the corresponding time differences. Coincidence events are then defined by applying timing cuts on these differences. In particular, double coincidences between the entrance and stop scintillators are used to identify particles traversing the setup, while additional timing requirements with respect to the MCP signals define triple-coincidence events associated with secondary electron emission.

\subsection{\label{sec:EffExtraction}Efficiency Extraction}

Particles traversing the full setup are identified by requiring coincident signals in the entrance and stop scintillators. The coincidence window is defined as a \SI{50}{\nano\second} interval after the entrance signal, determined from the measured time-of-flight distributions between the detector signals. The number of such events, denoted $N_{\mathrm{double}}$, serves as the normalization for the efficiency extraction. The contribution from accidental coincidences was evaluated using low-momentum datasets (e.g. \SI{8}{\mega\electronvolt\per c} and \SI{10}{\mega\electronvolt\per c}), for which no muon signal is expected in the stop detector, as muons with these momenta are stopped in the entrance detector or the tagging foil. No coincident events were observed over acquisition times of several tens of minutes. The accidental coincidence rate was therefore considered negligible. Consequently, no explicit background subtraction was applied.

As muons traverse the foil, secondary electrons may be emitted and detected by the two MCP detectors, which are treated independently in the analysis. For each MCP, triple-coincidence events are defined by requiring temporal coincidence with the two scintillator signals. Specifically, the time difference between MCP and the entrance detector signal is required to be within a \SI{25}{\nano\second} window, determined analogously from the corresponding time-of-flight distribution. If multiple MCP signals fall within the coincidence window of a given double-coincidence event, they are treated as a single detection, reflecting the passage of one particle through the setup.

The detection efficiency $\epsilon^\mathrm{meas}$ of the foil-based system is defined as the fraction of double-coincidence events associated with at least one detected secondary electron,
\begin{equation}
\epsilon^\mathrm{meas} = \frac{N_{\mathrm{matched}}}{N_{\mathrm{double}}},
\end{equation}
where $N_{\mathrm{matched}}$ denotes the subset of double-coincidence events that are associated with at least one MCP signal within the defined coincidence window. Efficiencies are extracted separately for each MCP, as well as for their logical combination (i.e.\ detection in either the backward or forward MCP, or in both). The statistical uncertainty on $\epsilon^\mathrm{meas}$ is evaluated assuming binomial statistics.

\begin{figure}[t!]
\centering
\includegraphics[width=1\columnwidth, trim={0 0 0 0},clip]{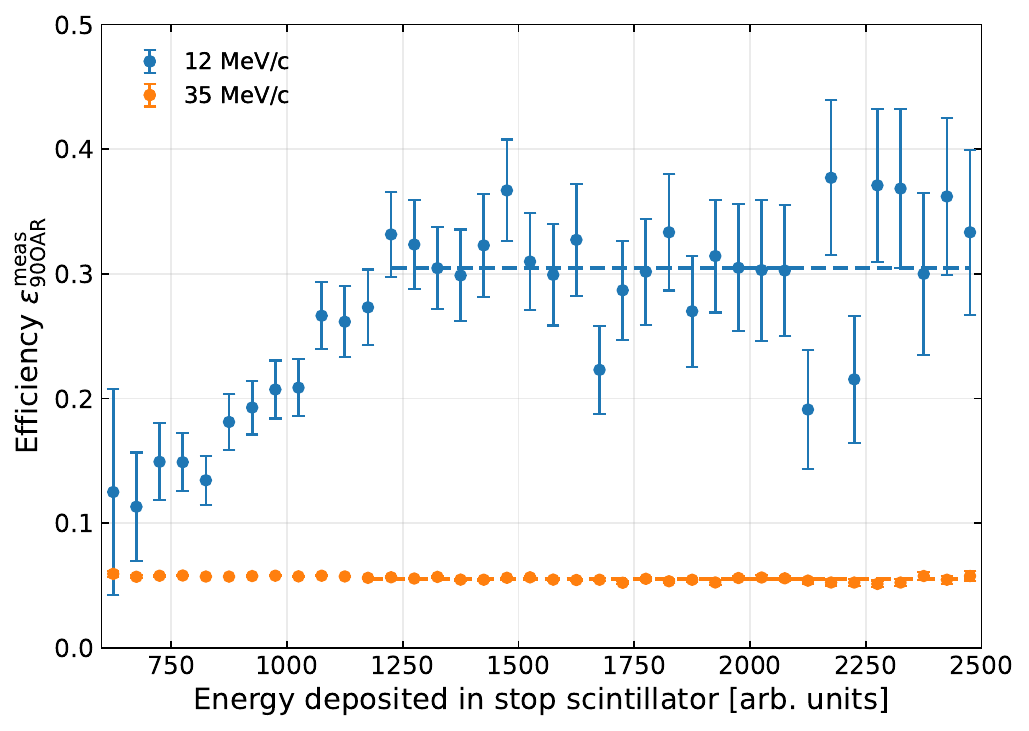}
\caption{\label{fig:cutscan}Detection efficiency $\epsilon_\mathrm{90OAR}^\mathrm{meas}$ as a function of the energy deposited in the downstream scintillator for beam momenta of \SI{12}{\mega\electronvolt\per c} and \SI{35}{\mega\electronvolt\per c}. At \SI{12}{\mega\electronvolt\per c}, a pronounced rise in efficiency is observed at low deposited energies, followed by a plateau at higher energies, indicating the transition from background-dominated events to muons traversing the full setup. The dashed line indicates the plateau value used for the efficiency extraction. In contrast, the \SI{35}{\mega\electronvolt\per c} data exhibit no significant dependence on deposited energy, consistent with negligible background contamination.}
\end{figure}

At lower momenta, an additional background contribution arises from muons stopping in the foil or nearby material. Because a fraction of these stopped muons decay shortly after stopping, their decay electrons can reach the downstream scintillator within the coincidence window and contaminate the double-coincidence sample. To suppress this contribution, the tagging efficiency was evaluated as a function of the energy deposited in the downstream scintillator. Events with low deposited energy exhibit an efficiency approximately three times smaller than that observed at higher deposited energies, indicating a dominant background contribution. In contrast, a stable plateau is observed at larger deposited energies, corresponding to events dominated by muons traversing the full setup. The final efficiency is therefore extracted exclusively from this plateau region. The plateau is identified by scanning contiguous energy intervals and selecting regions consistent with a constant efficiency within statistical uncertainties. This behaviour is most pronounced for the \SI{12}{\mega\electronvolt\per c} dataset and remains visible at \SI{16}{\mega\electronvolt\per c}, while no significant dependence on deposited energy is observed at higher momenta. A comparison of the energy-dependent efficiency for \SI{12}{\mega\electronvolt\per c} and \SI{35}{\mega\electronvolt\per c} is shown in Fig.~\ref{fig:cutscan}.

\subsection{\label{sec:beamspot}Beamspots}

\begin{figure*}[t!]
\centering
\includegraphics[width=2\columnwidth, trim={0 0 0 0},clip]{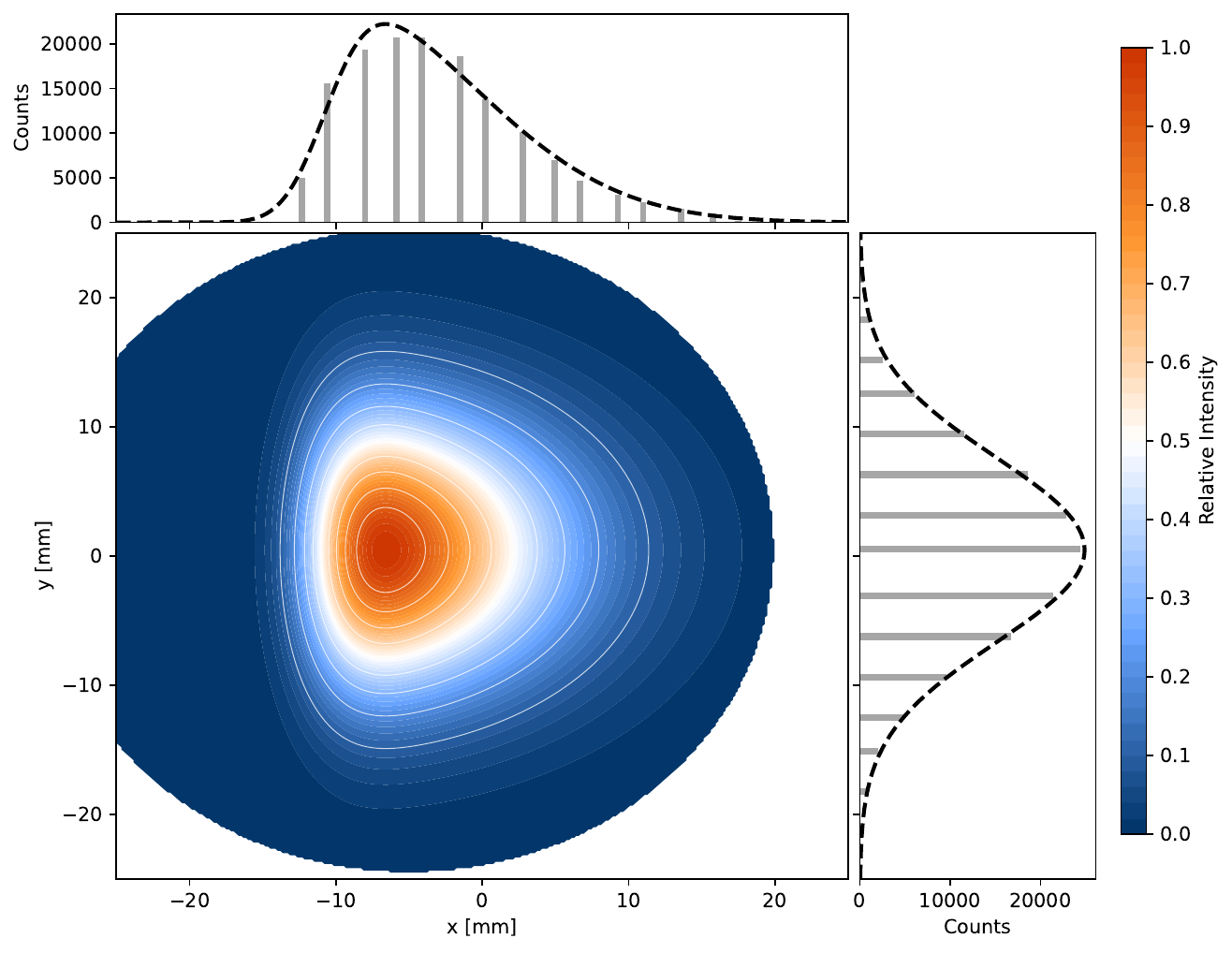}
\caption{\label{fig:beamspot}
Reconstructed transverse beam profile for \SI{35}{\mega\electronvolt\per c} negative muons obtained from position-sensitive measurements with the forward MCP of secondary electrons emitted from the foil. The central panel shows the spatial distribution of reconstructed events, with the color scale indicating relative intensity. The top and right panels display the corresponding projections onto the $x$ and $y$ axes, respectively, together with fits using skewed Gaussian functions (dashed lines). The observed asymmetry in the horizontal direction is consistent with the asymmetric electrode and ground-plane geometry between foil and MCP. Despite the limited time resolution of the data acquisition system, the underlying beam structure is captured, demonstrating the feasibility of foil-based beam monitoring.
}
\end{figure*}

In addition to the efficiency measurements, the position-sensitive MCP detectors enable a proof-of-principle reconstruction of the transverse muon beam profile, shown in Fig.~\ref{fig:beamspot}. Coincident MCP events with valid delay-line signals are used to reconstruct the transverse hit distributions on the detectors. The delay-line coordinates are obtained from the differences of the corresponding signal arrival times and converted to spatial coordinates using the detector calibration constants.

Due to the limited time resolution of the data acquisition system, individual MCP hits cannot be reconstructed with high spatial resolution. Nevertheless, the underlying beam profile is clearly visible and can be described by skewed Gaussian distributions. The reconstructed distributions for the two MCPs are shown in Fig.~\ref{fig:beamspot}, while the corresponding projections are fitted using skewed Gaussian functions.

To relate the measured MCP hit distributions to the muon impact positions at the foil, SIMION simulations were performed, incorporating both the initial energy distribution and the angular emission distribution~\cite{2016_Allegrini} of the secondary electrons. However, the reconstruction remains highly sensitive to the initial beam conditions, in particular the true beam profile at the foil position. A fully calibrated mapping between the foil and MCP would therefore require dedicated mask measurements, which were not yet possible with the current experimental setup. At the present stage, the reconstructed beam profiles should therefore be regarded as approximate.

Nevertheless, the simulations reproduce characteristic features observed in the data, such as the skewed Gaussian shape in the horizontal direction. The simulations indicate that the asymmetric electrode and ground-plane geometry contributes to this asymmetry by preferentially deflecting emitted electrons toward one side of the detector.

Additionally, the reconstructed beamspots provide an empirical estimate of the electron transport efficiency between foil and MCP. Electrons emitted from regions that are transported outside the active detector area reduce the observed tagging efficiency. For typical beam sizes with Gaussian widths of approximately 10 mm, the simulations predict transport efficiencies of around 90\%. This correction, further discussed in Sec.~\ref{sec:eletransport}, therefore contributes to the characterization of the overall detector response.

This proof-of-principle measurement demonstrates that foil-based tagging can be used for minimally invasive beam monitoring. With improved readout capabilities and reduced material budgets, the method also shows strong potential for future tracking applications.

\section{\label{sec:Corr}Corrections}

\subsection{\label{sec:mom_loss}Momentum Loss}

At low momenta, the interactions of muons with the material budget that produce the background shown in Fig.~\ref{fig:cutscan} also reduce the effective muon momentum at the foil. To account for this, the momentum loss in the material budget, consisting of the entrance detector and the aluminized Mylar foil, is evaluated using a dedicated \textsc{Geant4} simulation based on the framework developed for the MIXE experiment. This simulation is used to convert the nominal beamline momentum to the effective muon momentum immediately before and after the foil. In particular, the incident momentum at the foil is used for backward-emitted electrons, while the residual momentum after traversing the foil is used for forward-emitted electrons.

\begin{figure}[t!]
\centering
\includegraphics[width=1\columnwidth, trim={0 0 0 0},clip]{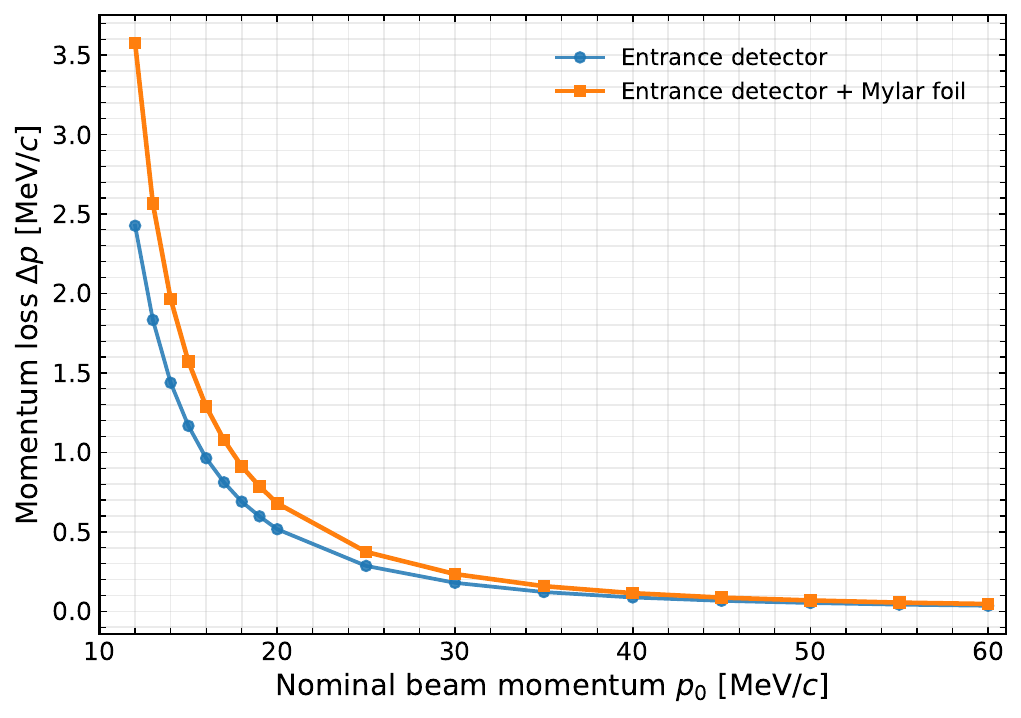}
\caption{\label{fig:ploss}Momentum loss $\Delta p$ as a function of the nominal beam momentum $p_0$, as obtained from \textsc{Geant4} simulations of the material budget. The contributions from the entrance detector alone and from the combined system of entrance detector and aluminized Mylar foil are shown separately. At low momenta, significant momentum loss is observed, dominated by the entrance detector, with an additional contribution from the foil. For higher momenta, the momentum loss rapidly decreases and becomes negligible above approximately \SI{30}{\mega\electronvolt\per c}.}
\end{figure}

The simulation shows that the momentum loss becomes significant only at the lowest beam momenta. In particular, the entrance detector leads to noticeable energy loss for \SI{12}{\mega\electronvolt\per c} and \SI{16}{\mega\electronvolt\per c}, while additional losses in the Mylar foil are only relevant at \SI{12}{\mega\electronvolt\per c}. For higher momenta, the momentum loss is negligible. The corresponding corrections are shown in Fig.~\ref{fig:ploss}.

\subsection{\label{sec:eletransport}Electron Transport Efficiency}

In addition to the intrinsic secondary electron emission probability and the MCP detection efficiency, the measured detector response depends on the transport efficiency of emitted electrons between foil and MCP. Electron losses are mainly caused by the finite active area of the MCP detector, such that broader beam spots lead to a larger fraction of transported electrons missing the detector. To estimate this contribution, the reconstructed beamspot distributions discussed in Sec.~\ref{sec:beamspot} are used.

\begin{figure}[t!]
\centering
\includegraphics[width=1\columnwidth, trim={0 0 0 0},clip]{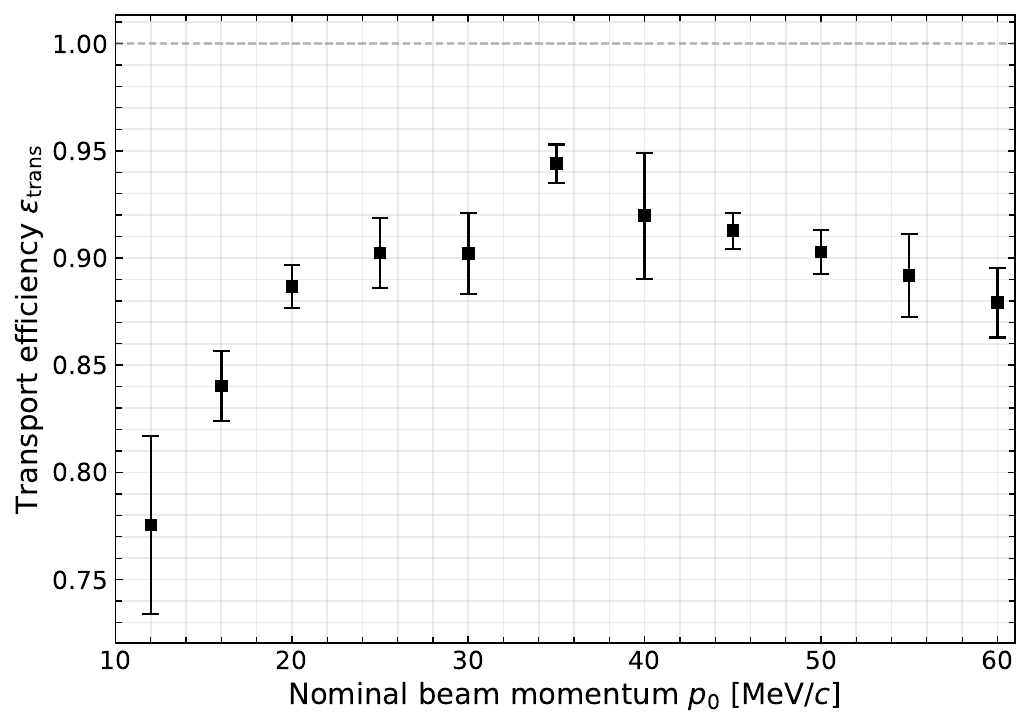}
\caption{\label{fig:transport_efficiency}
Electron transport efficiency $\epsilon_{\mathrm{trans}}$ between foil and MCP as a function of the nominal beam momentum $p_0$. The transport efficiency is estimated from the reconstructed beamspot distributions by determining the fraction of reconstructed events contained within the active MCP area. For most beam momenta, transport efficiencies around \SI{90}{\percent} are observed, indicating only minor acceptance losses during electron transport from foil to detector. At the lowest momenta, reduced transport efficiencies are observed due to the increased beamspot size caused by multiple scattering in the entrance detector material.}
\end{figure}

The transport efficiency $\epsilon_{\mathrm{trans}}$ is estimated from the fraction of reconstructed events contained within the active MCP area. The measured beam profiles are fitted using skewed Gaussian functions, which reproduce the characteristic asymmetries observed in the data. Integrating the fitted distributions over the active detector region allows the fraction of transported electrons reaching the MCP to be estimated. 
The computed transport efficiencies are shown in Fig.~\ref{fig:transport_efficiency} as a function of nominal beam momentum. The smallest transport efficiencies are observed at the lowest momenta, where the beamspot is strongly broadened by multiple scattering in the entrance detector.

Since the transport efficiency enters as an additional multiplicative contribution to the measured detector response, the measured efficiency needs to be corrected to
\begin{equation}
\epsilon_\mathrm{90OAR}^{\mathrm{corr}}=\frac{\epsilon_\mathrm{90OAR}^{\mathrm{meas}}}{\epsilon_\mathrm{trans}}
\end{equation}
when extracting the intrinsic secondary electron emission yield. Compared to the MCP open-area-ratio correction, the transport correction is smaller but introduces a non-negligible momentum-dependent contribution to the extracted SEY.

\section{\label{sec:rResults}Results and Discussion}

\begin{figure}[t!]
\centering
\includegraphics[width=1\columnwidth, trim={0 0 0 0},clip]{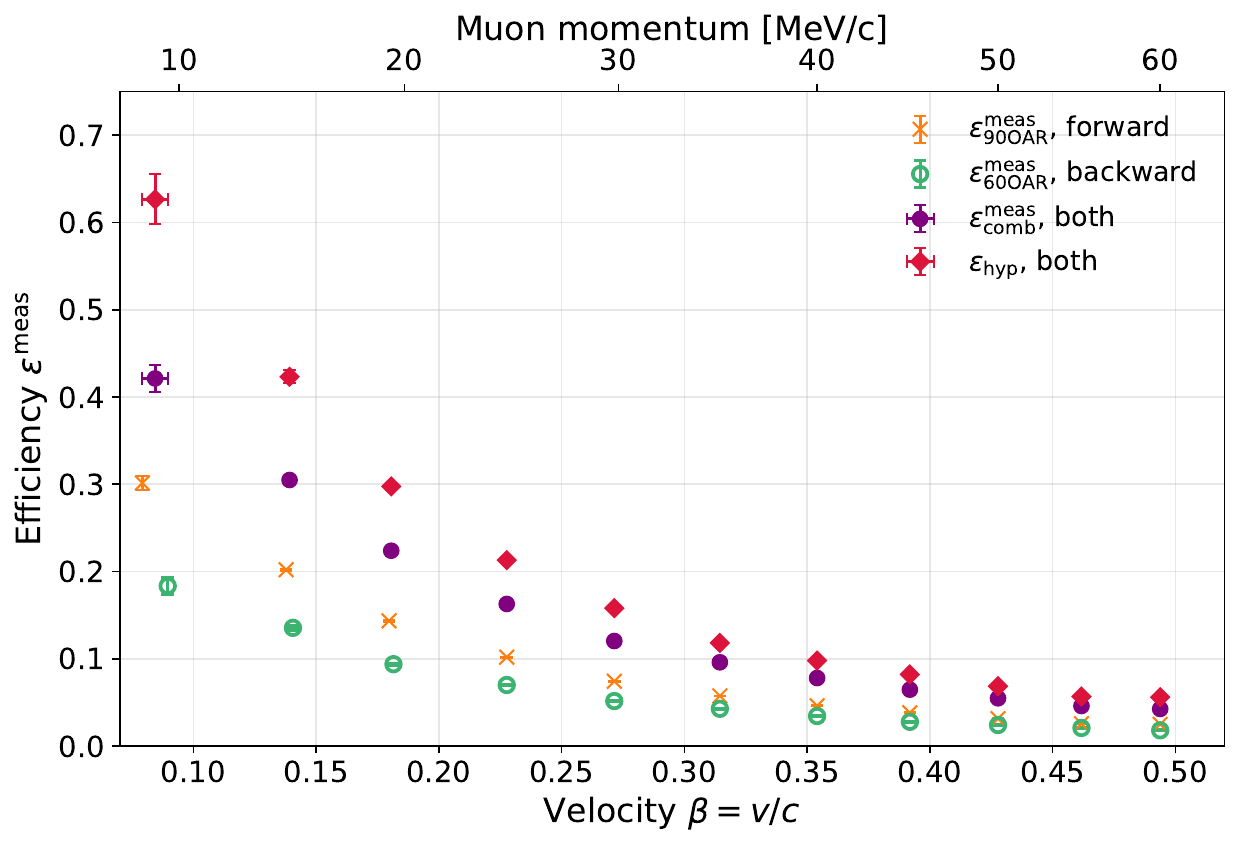}
\caption{\label{fig:exp_al_pie1}
Measured detection efficiency $\epsilon^\mathrm{meas}$ for muons traversing the aluminized Mylar foil as a function of the particle velocity $\beta = v/c$. The efficiency for detecting forward-emitted electrons with the 90\% OAR MCP ($\epsilon_{\mathrm{90OAR}}^\mathrm{meas}$, orange crosses) and backward-emitted electrons with the 60\% OAR MCP ($\epsilon_{\mathrm{60OAR}}^\mathrm{meas}$, green circles) are shown separately, together with their combined efficiency ($\epsilon_{\mathrm{comb}}^\mathrm{meas}$, purple circles). The red diamonds indicate the expected efficiency for a configuration with two identical 90\% OAR MCPs. For the combined and hypothetical efficiencies, the data points are plotted at the mean corrected momentum, while the horizontal error bars indicate the corrected muon momenta before and after traversing the foil, respectively. The corresponding muon momentum is shown on the top axis.}
\end{figure}

The measured detection efficiency $\epsilon^\mathrm{meas}$ for muons traversing the aluminized Mylar foil is shown in Fig.~\ref{fig:exp_al_pie1}. A clear dependence on muon momentum is observed, with increasing efficiency toward lower momenta. This behaviour reflects the rising SEY at lower particle velocities, consistent with previous experimental observations and theoretical models of ion-induced electron emission \cite{2019_Fazlul}.

The efficiencies measured for backward-emitted electrons ($\epsilon_\mathrm{60OAR}^\mathrm{meas}$, green circles) and forward-emitted electrons ($\epsilon_\mathrm{90OAR}^\mathrm{meas}$, orange crosses) differ systematically due to the different intrinsic detection efficiencies of the MCPs, which arise from their respective OARs. As expected, the forward detector exhibits a higher efficiency. The combined efficiency ($\epsilon_\mathrm{comb}^\mathrm{meas}$, purple circles), obtained by detecting electrons on both sides of the foil, significantly enhances the overall tagging performance.

To estimate the achievable performance for an optimized configuration, a hypothetical setup with two identical MCPs of 90\% OAR and perfect transport efficiency is considered (red diamonds in Fig.~\ref{fig:exp_al_pie1}). The corresponding efficiency is given by

\begin{equation}
\label{eq:hyp}
\epsilon_\mathrm{hyp}
=
1 -
\left(
1-
\epsilon_\mathrm{90OAR}^{\mathrm{corr}}
\right)^2,
\end{equation}

representing the probability that at least one of two identical detectors registers a secondary electron. This estimate indicates that substantially higher tagging efficiencies can be achieved with improved detector performance. 

\begin{figure}[t!]
\centering
\includegraphics[width=1\columnwidth, trim={0 0 0 0},clip]{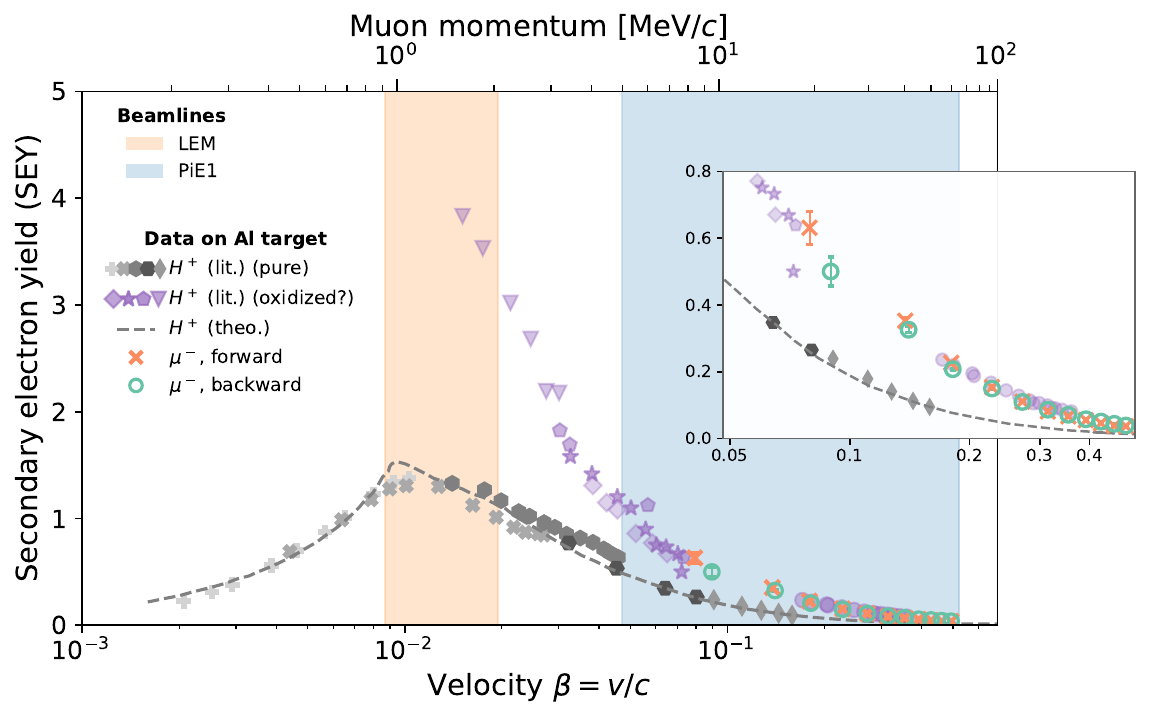}
\caption{\label{fig:SEY_aluminium}
Secondary electron yield for muons on aluminized Mylar as a function of particle velocity $\beta = v/c$. The data extracted in this work are shown for forward- and backward-emitted electrons, together with literature values for aluminium. Literature data corresponding to clean aluminium surfaces are shown in gray shades, while measurements without specific surface preparation (typically oxidized surfaces) are shown in purple shades. The dashed gray curve represents the theoretical model from Ref.~\cite{2019_Fazlul}. The shaded bands indicate the velocity ranges accessible at the LEM (orange) and $\pi$E1 (blue) beamlines. The inset shows a zoom of the $\pi$E1 region to highlight the measured data points.}
\end{figure}

To compare our measurements with literature data, the efficiencies are converted to secondary electron yield $\gamma$ using
\begin{equation}
\label{eq:SEY}
\gamma =-\frac{\ln\left(1-\epsilon^{\mathrm{corr}}\right)}{\eta_{\mathrm{MCP}}},
\end{equation}

where $\eta_{\mathrm{MCP}}$ denotes the single-electron detection efficiency of the MCP~\cite{2003_Allegrini_SEY}. Values of $\eta_{\mathrm{MCP}} = 0.54$ and $0.78$ are used for the backward and the forward MCP, respectively. The extracted SEY values are shown in Fig.~\ref{fig:SEY_aluminium} together with literature data for aluminium.

The literature data are grouped according to surface preparation, distinguishing clean aluminium surfaces \cite{1979_Baragiola_AlReview,1981_Svensson_AlReview,1981_Hasselkamp_AlReview,1995_Benka_AlReview,1981_Koyama_AlReview} from untreated or oxidized surfaces \cite{1954_Aarset_AlReview,1977_Thornton_AlReview,1974_Foti_AlReview,1939_Hill_AlReview,1997_Castanenda_AlO2_AlReview,1988_Borovsky}.
Previous studies~\cite{1997_Castanenda_AlO2_AlReview, 1988_Borovsky} have explicitly demonstrated that oxidized aluminium exhibits significantly higher SEY than clean aluminium. Since no special surface preparation was applied to the aluminized Mylar foil used in this work, the presence of oxide layers is expected. The measured data agree well with literature values for oxidized aluminium, supporting this interpretation and confirming the validity of the measurement.

In Fig.~\ref{fig:SEY_aluminium}, the accessible momentum ranges of representative beamlines are also indicated, including the LEM beamline (orange) and the $\pi$E1 beamline (blue) at PSI. The observed increase in SEY toward lower velocities implies that foil-based tagging becomes increasingly efficient in the low-momentum regime, which is particularly relevant for beamlines operating below approximately \SI{20}{\mega\electronvolt}/$c$.

\begin{figure}[t!]
\centering
\includegraphics[width=1\columnwidth, trim={0 0 0 0},clip]{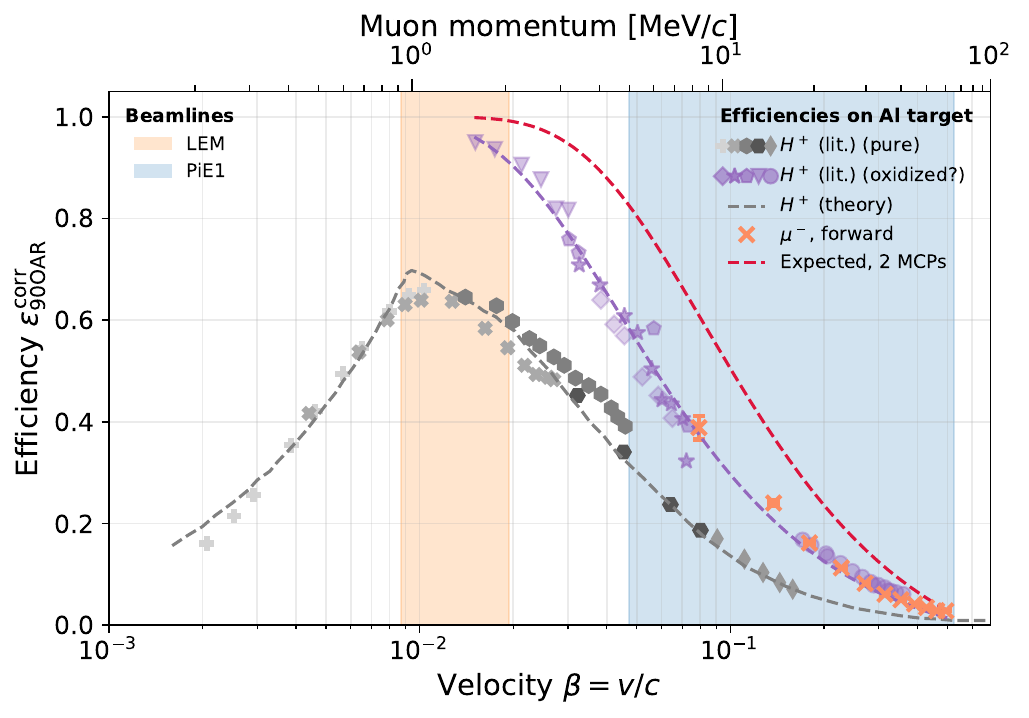}
\caption{\label{fig:eff_Al}
Detection efficiency $\epsilon_\mathrm{90OAR}^\mathrm{corr}$ as a function of particle velocity $\beta = v/c$ for aluminium targets. The corrected efficiencies for forward-emitted electrons (orange crosses) are shown together with efficiencies derived from literature SEY data for protons on aluminium. The latter are converted using Eq.~(\ref{eq:SEY}) assuming a single-electron detection efficiency corresponding to a 90\% OAR MCP. Literature data for clean aluminium surfaces are shown in gray shades, while oxidized or untreated surfaces are shown in purple shades. The dashed gray curve corresponds to the theoretical model from Ref.~\cite{2019_Fazlul}, converted to efficiency. The red dashed curve represents the expected efficiency for a configuration with two identical 90\% OAR MCPs detecting forward- and backward-emitted electrons, derived from the literature trend for oxidized aluminium. The shaded bands indicate the velocity ranges accessible at the LEM (orange) and $\pi$E1 (blue) beamlines.}
\end{figure}

An alternative representation is shown in Fig.~\ref{fig:eff_Al}, where literature SEY data are converted into detection efficiencies for a 90\% OAR MCP by reversing Eq.~\ref{eq:SEY}. The oxidized aluminium data are described by a sigmoid function to guide the eye. This representation suggests that foil-based tagging remains highly efficient at low velocities. Applying Eq.~\ref{eq:hyp} to the efficiency obtained from the fitted oxidized-aluminium trend yields the expected efficiency for a configuration with two identical 90\% OAR MCPs, represented by the red dashed curve. This comparison illustrates the benefit of detecting both forward- and backward-emitted electrons with high-efficiency MCPs.

At very low momenta, however, the foil thickness ultimately becomes a limiting factor, as muons may stop within the foil before traversing it completely. Since secondary electron emission is a surface effect, the foil thickness can be reduced without compromising the emission process, thereby mitigating this limitation. Aluminium foils can routinely be produced with thicknesses of the order of \SI{1}{\micro\meter}, but even thinner materials are desirable to minimize energy loss, multiple scattering, and straggling.

\begin{figure}[t!]
\centering
\includegraphics[width=1\columnwidth, trim={0 0 0 0},clip]{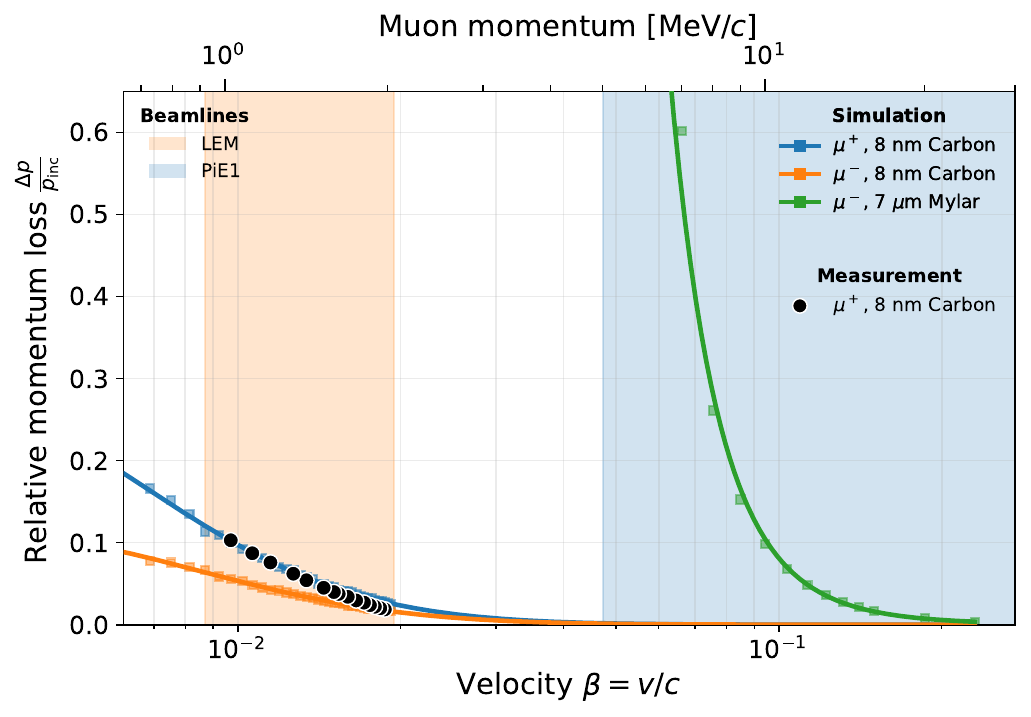}
\caption{\label{fig:sim_momentumloss}
Relative momentum loss $\Delta p / p_\mathrm{inc}$ as a function of particle velocity $\beta = v/c$ for different foil materials. Simulated results are shown for an \SI{8}{\nano\meter} amorphous carbon foil for positive muons ($\mu^+$, blue) and negative muons ($\mu^-$, orange), illustrating the charge-sign dependence of the stopping power due to the Barkas effect at low velocities. For comparison, the momentum loss in a \SI{7}{\micro\meter} Mylar foil (green) is shown, where the difference between $\mu^+$ and $\mu^-$ is negligible on the displayed scale. The shaded regions indicate the velocity ranges accessible at the LEM (orange) and $\pi$E1 (blue) beamlines. Black points represent measured momentum loss for positive muons on the carbon foil, demonstrating good agreement with the simulation. The strong increase in momentum loss for the Mylar foil at low velocities highlights the limitations of thicker materials, while ultrathin carbon foils introduce significantly smaller perturbations to the beam.}
\end{figure}
\begin{figure}[t!]
\centering
\includegraphics[width=1\columnwidth, trim={0 0 0 0},clip]{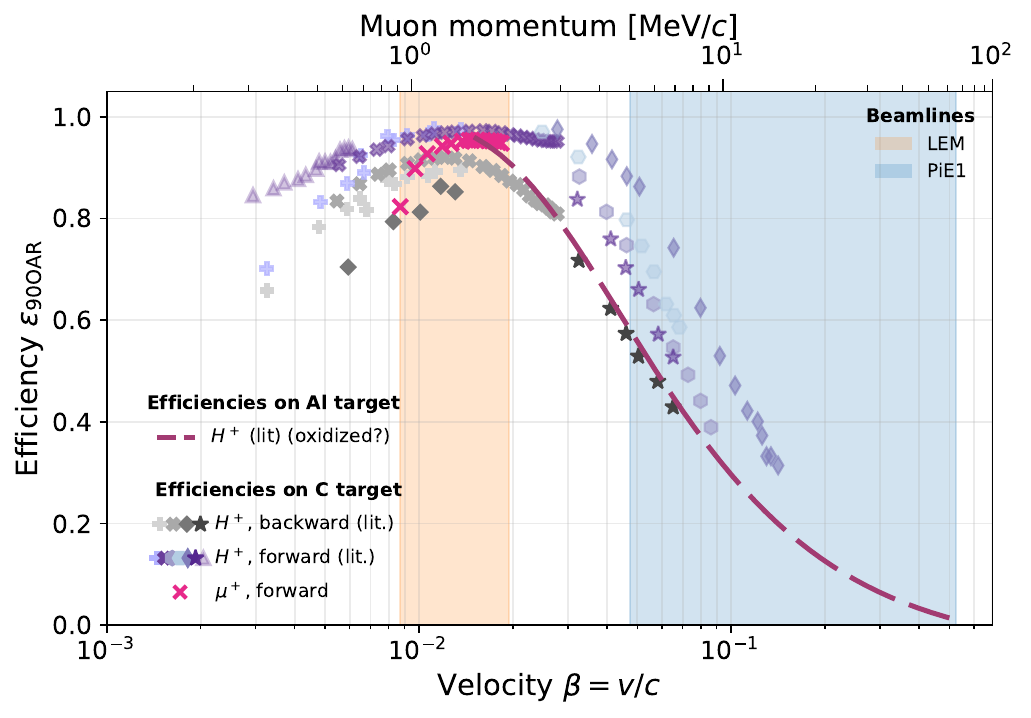}
\caption{\label{fig:eff_carbon}
Detection efficiency $\epsilon_\mathrm{90OAR}$ as a function of particle velocity $\beta = v/c$ for carbon targets. Efficiencies derived from literature SEY data for protons and muons on carbon are shown, converted using Eq.~(\ref{eq:SEY}) assuming a 90\% OAR MCP. For comparison, the red dashed curve represents the efficiency obtained from literature for oxidized aluminium, serving as a reference. The shaded regions indicate the velocity ranges accessible at the LEM (orange) and $\pi$E1 (blue) beamlines. The comparison suggests that ultrathin carbon foils can achieve efficiencies comparable to or exceeding those of aluminium while introducing significantly less perturbation to the beam.}
\end{figure}

To overcome this limitation, a promising alternative is the use of ultrathin carbon foils, with thicknesses on the order of \SI{10}{\nano\meter} or below. The impact of the material budget on the muon beam is illustrated in Fig.~\ref{fig:sim_momentumloss}, where the relative momentum loss $\Delta p / p_\mathrm{inc}$ is shown for a \SI{7}{\micro\meter} Mylar foil and an \SI{8}{\nano\meter} amorphous carbon foil. Here, $\Delta p$ denotes the momentum loss in the material and $p_\mathrm{inc}$ the incident momentum before any interaction.
For negative muon momenta in the range \SIrange{7}{30}{\mega\electronvolt\per c}, the simulation framework benchmarked by the MIXE experiment was used, while at lower momenta the \textsc{musrSim} package \cite{musrSIM1, musrSIM2, sedlak_musrsim_2012} was employed. In addition, low-energy positive muons were simulated and compared to experimental data (black points) from the LEM beamline~\cite{2024_Janka}. In this energy regime, the Barkas effect~\cite{1985_Ziegler} becomes significant, resulting in different stopping powers for positive and negative muons.

The simulations demonstrate that, in terms of energy loss and overall interaction strength, carbon foils introduce significantly less perturbation to the beam than Mylar foils, particularly at low momenta. This makes ultrathin carbon foils a more suitable choice for extending foil-based tagging into the lowest momentum regime.

To assess their tagging performance, literature data for protons and muons on carbon are shown in Fig.~\ref{fig:eff_carbon}. For comparison, the aluminium-based efficiency trend is included as a visual guide. The data indicate that carbon exhibits comparable or even higher SEY than (oxidized) aluminium, suggesting that such foils could enable efficient tagging across the full momentum range while introducing minimal disturbance to the beam.

\section{\label{sec:summary}Summary and Prospects}

In this work, the feasibility of foil-based muon tagging has been investigated for muon momenta between \SI{12}{\mega\electronvolt\per c} and \SI{60}{\mega\electronvolt\per c}. The measurements provide an experimental reference point in the intermediate-momentum regime, where conventional scintillator-based tagging becomes increasingly invasive at lower energies. By benchmarking the extracted secondary electron yields against literature data and models, the expected performance of foil-based tagging can be extrapolated toward lower-momentum muon regimes, thereby connecting the present measurements to ultrathin-foil techniques employed at \si{\kilo\electronvolt} energies.

To establish this connection experimentally, the secondary electron emission from a \SI{7}{\micro\meter} Mylar foil coated with aluminium was measured over the investigated momentum range. The results demonstrate that the emitted electron yield is sufficient to enable efficient foil-based particle tagging. The extracted detection efficiencies show a clear increase toward lower particle velocities, consistent with previous experimental observations and theoretical models from ion-induced electron emission processes. The measured SEY values are in good agreement with literature data for protons on aluminium, particularly for oxidized surfaces, supporting the validity of the measurement and the applicability of existing models to muons. In addition, the measurements provide new experimental data on muon-induced secondary electron emission in a previously unexplored momentum regime.

The achievable tagging efficiency is shown to depend strongly on the detector configuration. While the measured efficiencies are limited by the open-area ratio of the employed MCP detectors, an optimized setup using two high-efficiency (\SI{90}{\percent} OAR) MCPs would significantly enhance the overall detection probability, enabling efficiencies exceeding 50\% in the relevant velocity range.

At momenta below those investigated here, the foil thickness becomes a limiting factor, as muons may stop within the material before traversing it completely. This effect, observed for \SI{12}{\mega\electronvolt\per c} muons, highlights that the applicability of the method is ultimately constrained by the material budget rather than the SEY itself. Simulation studies suggest that ultrathin carbon foils, with thicknesses on the order of a few nanometers, drastically reduce energy loss and beam perturbation compared to Mylar foils, while maintaining comparable or higher SEY. This identifies carbon foils as a promising route to extend foil-based tagging into the lowest momentum regime.

In addition to timing capabilities, a proof-of-principle reconstruction of the muon beam profile has been demonstrated by correlating detected electron positions with their emission points at the foil. Despite limitations imposed by the current data acquisition system, the reconstructed beam profiles capture the main features of the underlying beam structure. This result paves the way for future foil-based detectors not only for particle tagging but also for minimally invasive beam monitoring.

Future developments will focus on improving the time and spatial resolution of the detection system, as well as on refining the electric field simulations governing electron transport. In combination with ultrathin target materials and optimized detector geometries, this approach offers a scalable solution for combined timing, beam diagnostics, and potentially tracking applications in low- and intermediate-energy muon beams.


\begin{acknowledgments}
All the measurements have been performed at the Swiss Muon Source S$\mu$S, Paul Scherrer Institute, Villigen, Switzerland. This work is funded by the Swiss National Science Foundation under the grant number 220823 (GJ). The authors would like to thank Andreas Suter for valuable discussions and technical support during the measurement campaign, and Paolo Crivelli for stimulating discussions that inspired the initial concept of this work.
\end{acknowledgments}


\bibliography{apssamp}
\end{document}